\documentclass[aip,amsmath,amssymb,reprint]{revtex4-1}

\usepackage{graphicx}
\usepackage{dcolumn}
\usepackage{bm}
\usepackage{filecontents}
\usepackage[utf8]{inputenc}
\usepackage[T1]{fontenc}

\begin{document}


\title{Temperature dependence of the magnetization of La$_{0.67}$Sr$_{0.33}$MnO$_3$ thin films on LaAlO$_3$}

\author{Arjan A. Burema}\altaffiliation{A. A. B and J. J. L. vR contributed equally to this work.}
\affiliation{University of Groningen, Zernike Institute for Advanced Materials,
9747 AG Groningen, The Netherlands}
\email{a.a.burema@rug.nl}

\author{Job J.L. van Rijn}\altaffiliation{A. A. B and J. J. L. vR contributed equally to this work.}
 \affiliation{University of Groningen, Zernike Institute for Advanced Materials,
9747 AG Groningen, The Netherlands}

\author{Tamalika Banerjee}
 \affiliation{University of Groningen, Zernike Institute for Advanced Materials,
9747 AG Groningen, The Netherlands}

\date{\today}

\begin{abstract}
We report on the interplay between magnetically ordered phases with temperature and magnetic field across compressively strained interfaces of thin La$_{0.67}$Sr$_{0.33}$MnO$_3$ films on LaAlO$_3$ substrates.   From the temperature dependence of the magnetization and resistivity studies, we find two distinct temperature regimes, where this interplay is clearly exhibited. We ascribe this to the strain induced Jahn-Teller like distortion that favors the stabilization of the d$_{3z^2-r^2}$ orbitals and enhances superexchange between adjoining Mn atoms. The temperature and field sweep of the magnetization and electronic transport leads to a hybridization between the closely spaced energy levels of d$_{3z^2-r^2}$ and d$_{x^2-y^2}$ orbitals leading to the coexistence of ferromagnetic and antiferromagnetic phases. Such an observation, not reported earlier, offer new routes for the design and study of magnetic textures in variously strained interfaces between perovskite oxides. 

\end{abstract}

\maketitle


\section{\label{sec:Introduction}Introduction}
Substrate engineering at perovskite heterointerfaces through strain, termination control or by oxygen octahedral coupling have emerged as promising routes to stabilize novel physical properties in these materials.\cite{dho_strain-induced_2003,ahn2004strain,houwman_out--plane_2008,boschker_uniaxial_2011,naftalis_out_2014,gabor_temperature_2015,roy_engineering_2016,feng_insulating_2016,liao2016controlled,liao2017experimental,huijben2017interface} For example, it has been shown that the induced epitaxial strain arising at different substrate interfaces with thin films of the same magnetic oxide can lead to the occurrence of diverse magnetic phases.\cite{wang2013oxygen} The epitaxial strain induced distortion of the oxygen octahedron in ABO$_3$ perovskite thin films lead to anisotropic hopping between different BO$_6$ orbitals and thus to differences in magnetic ordering in and out-of-plane of the film.\cite{nanda_effects_2008} For a compressively strained interface (a~<~c), an out-of plane ferromagnetic ordering and an in-plane antiferromagnetic ordering is usually favored, while the reverse is true for a tensile strained interface, whereas for an unstrained interface (a~$\sim$~c), a three dimensional ferromagnetic ordering can be stabilized. Lattice distortions at engineered interfaces are thus strongly coupled with magnetic properties and can lead to an anisotropy of magnetization along different crystalline directions or to a distribution of different magnetically ordered phases in magnetic films.
    
La$_\text{0.67}$Sr$_\text{0.33}$MnO$_3$ (LSMO) is a canonical ferromagnetic oxide in the manganite family whose application potential is driven by unique physical properties as well as its relatively high Curie temperature.\cite{tokura2000orbital} Besides its large colossal magnetoresistance and demonstration of 100\% spin-polarized carriers,\cite{park_direct_1998,urushibara1995insulator,Burgy_1,goodenough1997jb,de1997evidence,uehara1999percolative,bowen_nearly_2003,teresa_role_1999} it has also been used in different electronic and spintronic devices such as in magnetic tunnel junctions, hot-electron based transistors\cite{rana_hot_2013} as well as in ferroelectric tunnel junctions.\cite{garcia_ferroelectric_2014} The rich phase diagram of LSMO and the strong coupling between magnetic and electronic phases in thin LSMO films on different substrates \cite{adamo2009effect} such as SrTiO$_3$, \cite{boschker_uniaxial_2011,yin_strain-induced_2016,houwman_out--plane_2008} NdGaO$_3$,\cite{mathews_magnetization_2010} DyScO$_3$\cite{wang2013oxygen} and (LaAlO$_3$)$_{0.3}$-(SrAl$_{0.5}$Ta$_{0.5}$O$_3$))$_{0.7}$ (LSAT)\cite{boschker_uniaxial_2011} has been widely investigated. However, less is known about the coupling between magnetic and electronic properties in strained thin films of LSMO on a twinned LaAlO$_3$ (LAO) (100) oriented substrate. In this work, we study the interplay between coexisting magnetically ordered phases with temperature and magnetic field across compressively strained interfaces of thin LSMO films on LAO. We employ magnetization studies and electronic transport for this and find such an interplay to be active at different temperature regimes. These findings are further corroborated with the temperature dependence of the resistivity with and without an applied magnetic field across the heterointerface. Our findings indicate that the compressive strain rendered at such film-substrate interfaces lead to the stabilization of the d$_{3z^2-r^2}$ orbitals and to the coexistence of ferromagnetic and antiferromagnetic phases. The latter is a consequence of the stretching of Mn-O-Mn bond lengths and bond angles at the interface, altering the overlap between the MnO$_6$ orbitals. This coupling between the spin and orbital degrees of freedom at such engineered interfaces offers interesting prospects for the design and study of new magnetic textures at such interfaces. 

\section{\label{sec:Fabrication}Fabrication and structural characterization}

\subsection{\label{sec:Thinfilmgrowth}Thin film growth}
\begin{figure}[h!]
    \centering
    \includegraphics[width=\linewidth]{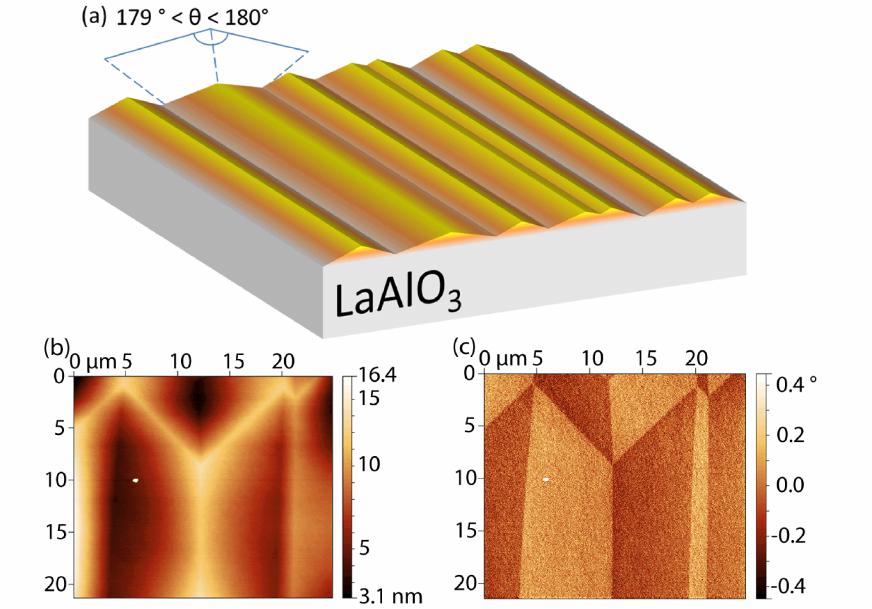}
        \caption{(color online) (a) A schematic of the top surface of a twinned LAO substrate (not to scale). The angle between the two faces of a typical twin boundary is indicated. (b) The AFM image shows twinning of a LAO substrate. The twinning varies in width between 1 $\mu$m and 7 $\mu$m and is tens to hundreds of microns long. The height of a twin is between 5 nm and 20 nm, hence three orders of magnitude smaller compared to the width. The resulting tilting angle of two faces on either side of the twin boundary is between 179~$ \mathring{ }$ and 180~$ \mathring{ }$. (c) The phase image of the figure clearly shows the contrast of the twin faces on either side of the boundary.}
    \label{fig:AFM}
\end{figure}

The 15 unit cell (u.c.) thick LSMO film is grown using Pulsed Laser Deposition (PLD) on a 5x5 mm LAO substrate. Optical microscope image shows a twinned surface, typical of a LAO substrate. Additionally, the twinned surface of LAO is visualized by Atomic Force Microscopy (AFM), as shown in Fig. \ref{fig:AFM}. From these observations, we find the twinning width to be several microns whereas the length ranges from tens to hundreds of microns. The highest and lowest point of the twinned plane is found to be up to 20 nm in height. The size of the twins is smaller than the Hall bar (100 by 3000 $\mu$m) and the Hall bar includes tens to hundreds of twins. 

The depositions were done in a Twente Solid State Technology PLD system with an excimer (KrF) laser (248 nm), at 750~$^{\text{o}}$C under a pure oxygen pressure of 35~Pa. A laser fluence of 2~J/cm$^2$ at a frequency of 1 Hz is used with an oxygen flow of 1~cm$^3$/s and a pressure of 35~Pa in the PLD chamber. The deposition is monitored using an \textit{in situ} Reflection High-Energy Electron Diffraction (RHEED) system. The RHEED spots are captured every 795 milliseconds. Figure \ref{fig:XRD}(a) shows post-deposition processing of the RHEED images for the analysis of the RHEED oscillations. This analysis provides information on the thickness of the film, considering one RHEED oscillation to correspond to the growth of one monolayer of LSMO. From the RHEED image, 15 oscillations are seen, establishing that a 15 u.c. thick LSMO film is grown on top of LAO. After deposition, the sample is cooled down at a rate of 10~$^{\text{o}}$C/min to room temperature, under an oxygen pressure of 1~kPa in order to anneal the film. The post-deposition annealing of the LSMO minimizes oxygen vacancy related defects. 

\subsection{\label{sec:Thinfilmcharacterization}Thin film characterization}
\begin{figure}[h!]
    \centering
    \includegraphics[width=\linewidth]{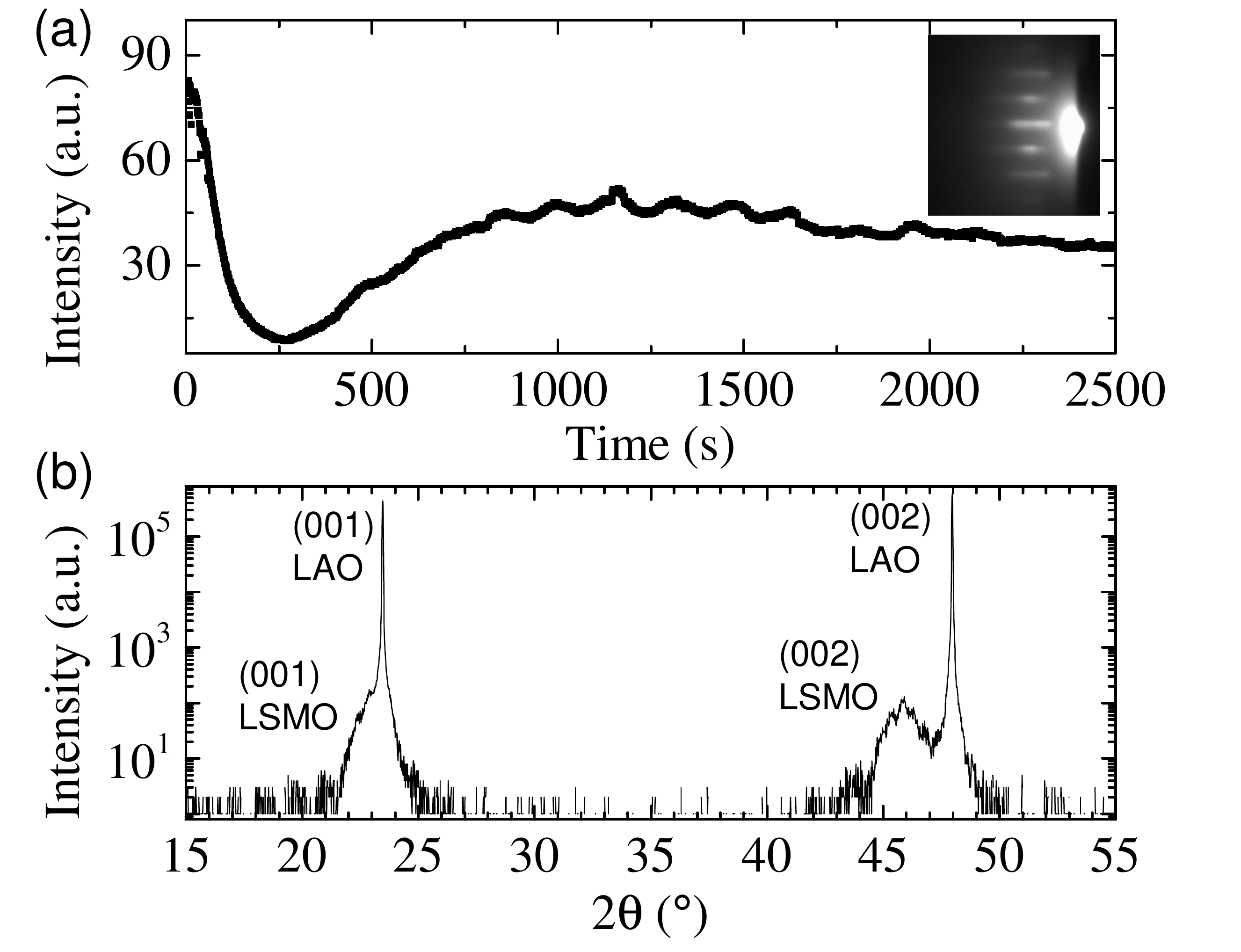}
    \caption{(a) The RHEED intensity oscillations of the first diffraction spot. The inset shows the RHEED spots, 5 min after the deposition of LSMO and is used to obtain the RHEED oscillations. The displayed intensity oscillations originate from the central spot. Analysing the RHEED intensity, 15 oscillations are visible establishing that 15 u.c. of LSMO are deposited on the LAO substrate. (b) XRD of a LSMO thin film on LAO. The intensity peaks correspond to the (001) and (002) peaks from LSMO and LAO. The absence of peaks in other reciprocal crystal directions suggest epitaxial growth of the film. A 2.1\% out-of-plane lattice strain is determined, comparing the extracted lattice parameter of 0.396~nm with that of the bulk value of 0.388 nm.}
    \label{fig:XRD}
\end{figure}

The structural and magnetic properties of the film are characterized by X-Ray Diffraction (XRD) (PANalytical) and magnetic property measurement system (Quantum design) measurements. The XRD 2$\theta$ scan of the (001) and (002) peaks shows a 2.1\% out-of-plane tensile strained LSMO film on top of the LAO substrate as shown in Fig. \ref{fig:XRD}(b). We used the same figure (Fig 2b) and fitted the LSMO (002) Bragg peak and the fringe on the left side of the peak to confirm the thickness of 15 u.c.. In the XRD 2$\theta$ scan with $\omega$=0, only [00l] peaks are visible, suggesting that the film is grown epitaxially in the out-of-plane direction. \\

\begin{figure}[h!]
    \centering
    \includegraphics[width=\linewidth]{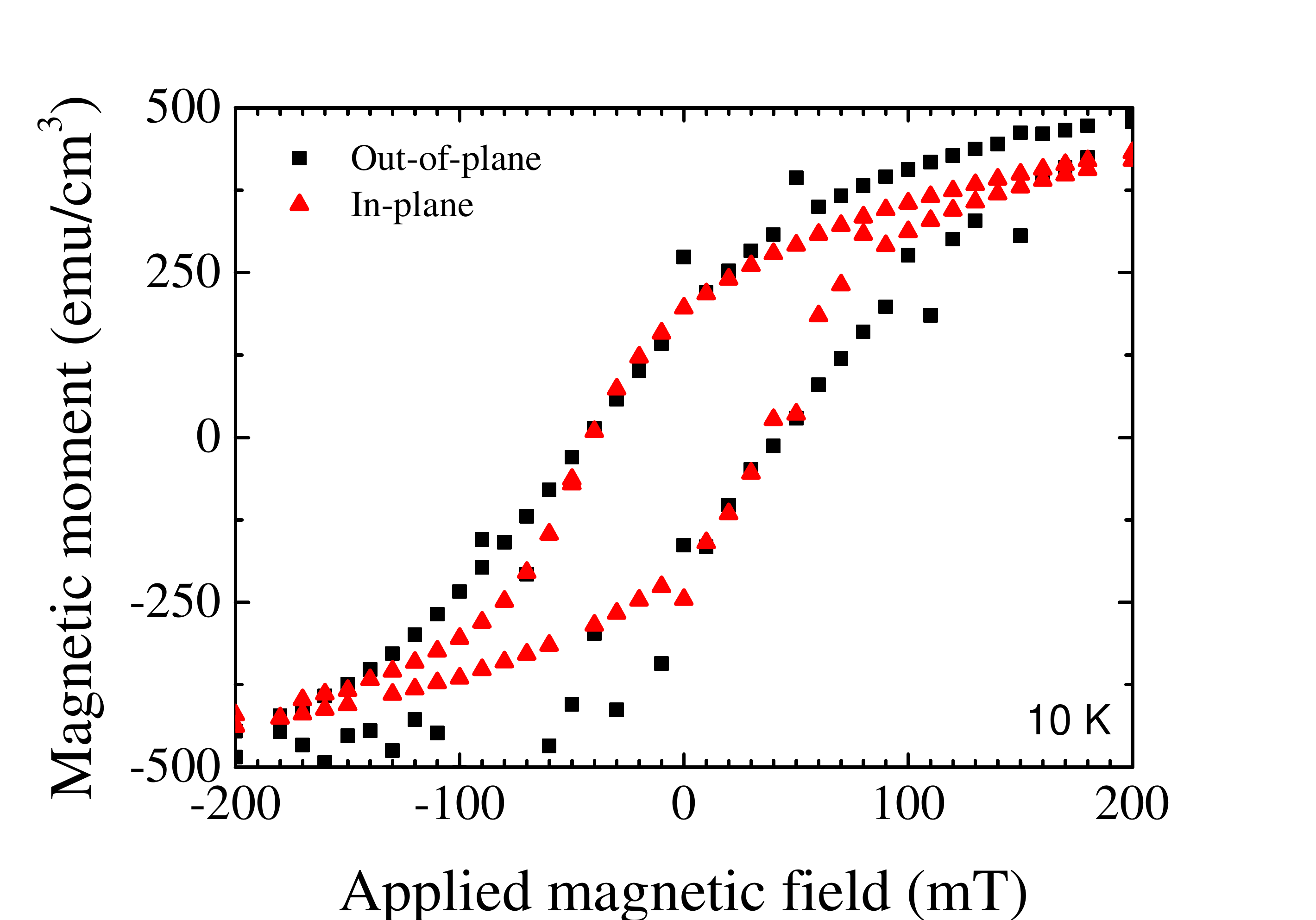}
    \caption{(color online) The magnetization with respect to the applied magnetic field for out-of-plane (square) and in-plane (up-triangles) field directions. The ferromagnetic loops at 10~K have similar coercive fields. Initially, a magnetic field of -2~T is applied, thereafter it is swept to 2~T and back to -2~T. Between the out-of-plane and in-plane measurements, the film is warmed above the Curie temperature. The saturation magnetization (M$_s$) of 570~emu/cm$^3$ is determined from the full loop (not shown).}
    \label{fig:MH}
\end{figure}

The magnetic property measurement system is used to perform magnetization measurements with temperature and applied magnetic field. In Fig. \ref{fig:MH}, the magnetization dependence on applied field is shown, where the magnetic field is swept both in-plane and out-of-plane with respect to the sample surface. The field dependent loops show ferromagnetic behavior. The coercive field of 43 mT at 10 K is similar for in-plane and out-of-plane measurements, indicating weak anisotropy. The highest saturation magnetization (M$_s$) is determined to be 570 emu/cm$^3$ at 2 T for 10 K. \\

\begin{figure}[h!]
    \centering
    \includegraphics[width=\linewidth]{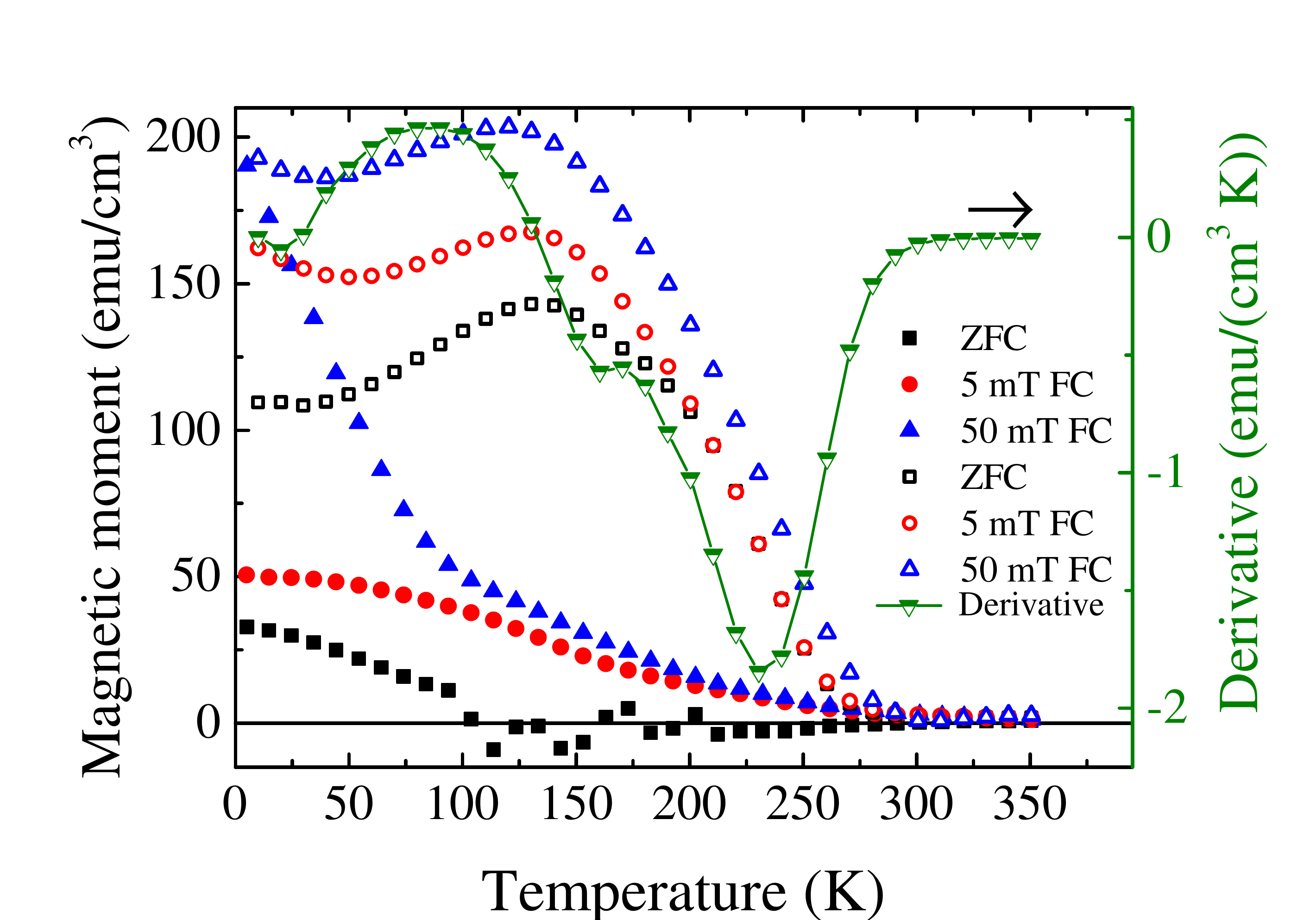}
    \caption{(color online) The magnetization, on the left axis, is measured during heating after cooling down to 10~K. The closed symbols correspond to the out-of-plane magnetization, while the open symbols correspond to the magnetization in-plane. A measuring field of 5~mT was applied. In addition to the ZFC measurement, FCW experiments were performed with cooling fields of 5~mT and 50~mT.  The Curie temperature, defined as the lowest point in the derivative, on the right axis, is at 230~K, which is lower than reported bulk values. The plotted derivative is taken from the ZFC in-plane measurement. A second phase transition is observed at 85~K. The out-of-plane (closed symbols) show decreasing magnetization with temperature and reaches a negligible value around 100 K. }
    \label{fig:MT}
\end{figure}

 The magnetization versus temperature, shown in Fig. \ref{fig:MT}, represents the Zero Field Cooling (ZFC) and two Field Cooled Warming (FCW) measurements. From the curves, two magnetic phase transition temperatures are extracted, around 230~K and 85~K. The maximum magnetization is reached around 130~K for the ZFC in-plane curve. Below 130~K the magnetization decreases for the in-plane component while the out-of-plane magnetization increases around 100~K. The in-plane magnetization deviates from a typical ferromagnetic behavior for LSMO films on LAO. The magnetization curves are an indication of an interplay between different magnetic phases in the film with temperature and the cooling field.

The in-plane magnetization shows a clear ferromagnetic phase between 130~K and 270~K. At lower temperatures, the magnetization decreases, suggesting the presence of an antiferromagnetic phase. The ZFC plots for the out-of-plane direction show the occurrence of spontaneous magnetic moments below 130~K. The 50~mT and 5~mT FCW out-of-plane magnetization plots show a similar trend between 130~K and 270~K, however, below 130~K the 50~mT FCW out-of-plane magnetization increases significantly. From the trends observed for ZFC plots at 130~K, we infer an interplay between coexisting ferromagnetic and antiferromagnetic phases. The 50~mT FCW measurements show that this cooling field is sufficient to order the existing ferromagnetic phase in the out-of-plane direction.

\section{\label{sec:Magnetotransport}Magnetotransport}
\begin{figure}[h!]
    \centering
    \includegraphics[width=\linewidth]{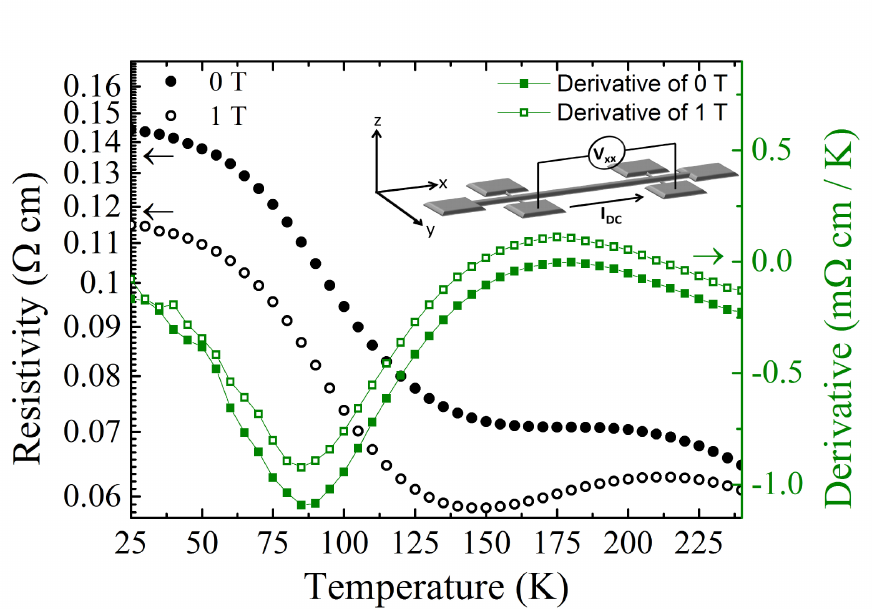}
    \caption{(color online) Temperature dependence of the resistivity, left axis, for the 15~u.c. LSMO film with and without a magnetic field (1 T) and with a direct current of 5~$\mu$A. 
    The temperature dependence of the derivative of the resistivity, with and without a magnetic field (1~T), is shown in the right axis. From the derivative plots two phase transition temperatures at 85~K and at 175~K are determined. (Inset) A schematic of the Hall bar with spatial definitions. The field is applied in the positive z-direction and current is applied in the positive x-direction.}
    \label{fig:resitivity}
\end{figure}

In addition to the bulk film properties, magnetotransport measurements are performed in a Hall bar geometry. The Hall bar geometry is created using UV-lithography followed by wet etching in aqua regia to etch LSMO, using an established protocol.\cite{Gaurav} A schematic of the Hall bar is shown in the inset of Fig. \ref{fig:resitivity}. 

For the longitudinal resistivity measurements, the field is either set to 0~T or 1~T, in the z-direction, and a direct current of 5~$\mu$A is applied. By sweeping the temperature from 25~K to 240~K and simultaneously measuring the longitudinal voltage, the temperature dependence of the longitudinal resistivity is determined and plotted in Fig.~\ref{fig:resitivity}. The resistivity measurements show an insulating trend below 130 K, which is expected for a 15~u.c. thin film LSMO \cite{dois:10.1063/1.4926922} on LAO.\cite{yin_strain-induced_2016} In the temperature range between 145~K and 210~K, both the ZFC and 1~T curve shows a regime where the conduction is metallic. This indicates that there is phase coexistence in the material below 130~K and corroborates with our findings from M-T measurements. \\

From Fig. \ref{fig:resitivity}, we observe two distinct resistivity regimes with temperature. For temperatures between 25~K to 85~K, the resistivity changes from weakly insulating to metallic, while above 85~K, the LSMO film resistivity changes from metallic to weakly insulating. The resistivity increase in the low temperature regime is reminiscent of an antiferromagnetic phase, which, as suggested from earlier works \cite{wang2013oxygen,aruta2006strain,adamo2009effect} can occur under different growth conditions of the LSMO film or for different choices of the underlying substrate.

At higher temperatures, the resistivity of LSMO, transiting from metallic to insulating corresponds to the commonly observed magnetic phase transition from a ferromagnetic metal to a paramagnetic insulating phase.\cite{Burgy_1,goodenough1997jb,de1997evidence,uehara1999percolative}

The applied magnetic field of 1 T is larger than the magnetic saturation field and is expected to align all spins that are in the ferromagnetic state. This will also lead to an overall decrease of the resistivity. This reflects the strong coupling between magnetic and electronic states in manganite thin films. This reflects the strong coupling between magnetic and electronic states in manganite thin films. The resistivity studies also matches well with the M-T findings discussed earlier. From the derivative plots two phase transition temperatures at 85 K and at 175 K are determined.

\section{\label{sec:Discussion}Discussion}

Both the temperature dependence of magnetization and resistivity shown in Fig.~\ref{fig:MT} and Fig.~\ref{fig:resitivity} respectively suggest phase coexistence in our thin films.\cite{ahn2004strain,wang2013oxygen} Phase coexistence in tensile strained LSMO films deposited on DyScSr$_{0.3}$ accompanied by a phase transition from ferromagnetic to A-type antiferromagnetic behavior with changing temperature was reported earlier.\cite{wang2013oxygen}  Moreover, earlier works utilizing x-ray absorption spectroscopy have reported on the observation of C-type antiferromagnetism in ultrathin films of LSMO on LAO and its insulating behavior.\cite{aruta2006strain} However, for compressively strained thin LSMO films on LAO substrates, such as in our case, a phase coexistence has not been reported earlier. Our findings show the interplay between ferromagnetic metal and an antiferromagnetic insulator with temperature and applied magnetic field. \

Epitaxial growth of strained LSMO thin films on LAO substrate causes a deformation of the pseudo-cubic structure, elongating the crystal in the [001] direction and compressing it in the [100] and [010] direction. The strain resulting in a Jahn-Teller-like distortion,\cite{jahn_teller_1937stability} favors the stabilization of the d$_{3z^2-r^2}$ orbital over d$_{x^2-y^2}$ orbitals for compressively strained films.\cite{tokura2000orbital} Since the Mn-O-Mn bond angles and bond lengths are modified at such strained interfaces, this alters the overlap between the MnO$_6$ orbitals. The favorable occupation of the d$_{3z^2-r^2}$ orbital rather than the d$_{x^2-y^2}$ orbital induces low hopping probability via double exchange and enhances superexchange between Mn atoms, favoring an antiferromagnetic orientation of the spins and this is captured in Fig.~\ref{fig:MT}. The dominant ferromagnetic phase observed in many thicker films of LSMO on LAO,\cite{tebano2006strain} originates from the well-described double exchange between the Mn atoms. We find from both Fig. \ref{fig:MT} and Fig. \ref{fig:resitivity}, the presence of insulating antiferromagnetic phase below 130 K and a dominant ferromagnetic phase between 130 K and 270 K. For such Jahn-Teller like distorted heterointerfaces, the thermal energy is sufficient to induce hybridization between d$_{3z^2-r^2}$ and d$_{x^2-y^2}$ orbitals due to their relatively close-spaced energy levels,  leading to the observation of the interplay between the two magnetically ordered phases with temperature and applied field.\\

\section{\label{sec:Conclusion}Conclusion}

Strain engineered interfaces of thin LSMO films on LAO were studied using magnetization and electronic transport measurements. Two distinct temperature regimes are observed that are associated with an antiferromagnetic and a ferromagnetic phase. Such a coexistence of different magnetically ordered phases in compressively strained interfaces arises due to the stabilization of the d$_{3z^2-r^2}$ orbital over d$_{x^2-y^2}$orbitals and has not been reported earlier for such interfaces. The possibility to tailor the spin and orbital degrees of freedom across variously strained interfaces can be extended to the study and stabilization of non collinear magnetic textures across perovskite oxide interfaces.

\section{\label{sec:Acknowledgements}Acknowledgements}

We thank A. Das, P. Zhang, A. S. Goossens and S. Chen for useful scientific discussions and J. G. Holstein and H. H. de Vries for technical assistance. This work was realized using NanoLab NL facilities and is a part of the research program Skyrmionics: towards new magnetic skyrmions and topological memory (project number 16SKYR04). A. A. B acknowledges financial support from the Netherlands Organisation for Scientific Research (NWO).


%

\end{document}